\documentclass[a4paper,11pt]{article}
\pdfoutput=1

\usepackage{amssymb,mathrsfs,amsmath}
\usepackage{a4wide}
\usepackage{color,xcolor}
\usepackage[table]{xcolor}
\usepackage{slashed,soul}
\usepackage{graphicx}
\usepackage{amsfonts}
\usepackage{cancel}
\usepackage{lscape}
\def\linkcolor{cyan!70!black}

\usepackage[
colorlinks=true
,urlcolor=\linkcolor
,anchorcolor=\linkcolor
,citecolor=\linkcolor
,filecolor=\linkcolor
,linkcolor=\linkcolor
,menucolor=\linkcolor
,linktocpage=true
,pdfproducer=medialab
,pdfa=true
]{hyperref}
\usepackage{amsthm}
\usepackage{booktabs}
\usepackage{array}
\usepackage{rotating}
\usepackage[numbers, sort&compress]{natbib}
\usepackage{multirow}
\usepackage{float}
\usepackage[utf8]{inputenc}
\usepackage[T1]{fontenc}
\usepackage{appendix}
\usepackage{orcidlink}
\usepackage{xfrac}
\usepackage{arydshln}
\usepackage{xfrac}
\usepackage{dsfont}
\usepackage[normalem]{ulem}
\usepackage{cleveref}

\renewcommand{\baselinestretch}{1.2}

\allowdisplaybreaks

\begin{document}

\vspace{1cm}

\begin{titlepage}

% \vspace*{-1.0truecm}
% \begin{flushright}
% #preprints
%  \end{flushright}
% \vspace{0.8truecm}

\begin{center}
\renewcommand{\baselinestretch}{1.8}\normalsize
\boldmath
{\LARGE\textbf{
Neutrino t-channels at Colliders: \\ When Light Neutrinos Matter
}}
\unboldmath
\end{center}

\vspace{.2cm}

\renewcommand*{\thefootnote}{\fnsymbol{footnote}}

\begin{center}
{
\bf
Claudia Garcia-Garcia$^{1,2}$\footnote{\href{claudia.garciagarcia@unibo.it}{claudia.garciagarcia@unibo.it}}\orcidlink{0000-0003-2843-4081}, 
Manuel Gonz\'alez-L\'opez$^{3}$\footnote{\href{mailto:manuel.gonzalez@universidadunie.com}{manuel.gonzalez@universidadunie.com}}\orcidlink{0000-0001-7276-2192},
Xabier Marcano$^{1,2}$\footnote{\href{mailto:xabier.marcano@unibo.it}{xabier.marcano@unibo.it}}\orcidlink{0000-0003-0033-0504},
Daniel Naredo-Tuero$^{4}$\footnote{\href{mailto:daniel.naredo@kit.edu}{daniel.naredo@kit.edu}}\orcidlink{0000-0002-5161-5895}
}
\vspace{0.3truecm}

{\footnotesize
$^1$Dipartimento di Fisica e Astronomia, Universit\`a di Bologna, via Irnerio 46, 40126 Bologna, Italy
\\[.5ex]
$^2$INFN, Sezione di Bologna, viale Berti Pichat 6/2, 40127 Bologna, Italy
\\[.5ex]
$^3$Escuela Superior de Ingenier\'ia, Ciencia y Tecnolog\'ia, UNIE Universidad, c/Arapiles 14, 28015 Madrid, Spain
\\[.5ex]
$^4$ Institute for Astroparticle Physics (IAP), Karlsruhe Institute of Technology (KIT), \\[-1.5ex]
Hermann-von-Helmholtz-Platz 1, 76344 Eggenstein-Leopoldshafen, Germany
}

\vspace*{5mm}
%\today
\end{center}

\renewcommand*{\thefootnote}{\arabic{footnote}}
\setcounter{footnote}{0}

\vspace{0.4cm}
\begin{abstract}
\noindent 

Heavy Neutral Lepton (HNL)-mediated t-channel processes provide a unique opportunity to probe mass scales beyond the kinematic reach of direct production at high-energy colliders. We revisit these processes using the vector boson scattering channel $WW\to\ell\ell$ at the LHC as a case study, highlighting the essential role of the light neutrinos in restoring the proper high-energy unitary behavior of the scattering amplitude. Their inclusion, overlooked in some previous studies, leads to destructive interference that strongly suppresses lepton number violating signatures, demonstrating that a consistent treatment of the full seesaw spectrum qualitatively alters the phenomenology of t-channel HNL searches. This motivates the exploration of lepton number conserving but lepton flavor violating final states instead. We present a detailed analysis of the $pp\to e\mu jj$ channel and show that it provides a promising probe of TeV-scale HNLs in low-scale seesaw scenarios with sizable active--sterile mixing, extending the LHC sensitivity beyond existing direct searches.

\end{abstract}

\end{titlepage}

\tableofcontents

 %%%%%%%%%%%%%%%%%%%%%%%%%%%%%%%%%%%%%%%%%%%%%%%%%%%%%%%%%%%%%%%%%
 \section{Introduction}
 \label{Sec:Intro}

The type-I seesaw mechanism~\cite{Minkowski:1977sc,Mohapatra:1979ia,Yanagida:1979as,Gell-Mann:1979vob} is our simplest and most elegant extension of the Standard Model (SM) providing an explanation for the smallness of neutrino masses. It only requires the introduction of new, massive right-handed neutrinos, also known as sterile neutrinos or heavy neutral leptons (HNLs)\footnote{Their name often depends on the context, community or even epoch. We will mostly use the HNL terminology, as it is the most extended one in current collider works.}, which could also be connected to other open problems in particle physics such as the nature of Dark Matter~\cite{Dodelson:1993je,Shi:1998km,Abazajian:2001nj,Asaka:2005an} or the origin of the matter-antimatter asymmetry via Leptogenesis~\cite{Mohapatra:1986bd,Fukugita:1986hr}. Consequently, HNLs are one of the best motivated Beyond the SM (BSM) particles. 

This model introduces a new scale, associated to the Majorana mass of the new fields, independent of any other scale in the SM and completely unknown from the theory side. For this reason, a strong experimental program covering a wide range of HNL masses has been, is, and will be carried out. 
We refer the reader to Ref.~\cite{Fernandez-Martinez:2023phj} for an updated repository of current laboratory bounds, while future perspectives can be found, for instance, in Ref.~\cite{Abdullahi:2022jlv}. Additional and complementary information can be obtained from cosmological and astrophysical observations, see for instance, Refs.~\cite{Vincent:2014rja,Abazajian:2017tcc,Boyarsky:2020dzc,Dasgupta:2021ies}.

In this work we are interested in the heavy mass regime, above the electroweak (EW) scale. In said regime, the strongest limits are currently set by their low-energy effects, which induce deviations from unitarity in the neutrino mixing matrix~\cite{Broncano:2002rw}. The latest analysis of flavor and EW precision observables within this context was performed by Ref.~\cite{Blennow:2023mqx}. While still far from probing the canonical type-I seesaw model, these provide strong bounds for the symmetry protected low-scale seesaw realizations, such as the Inverse~\cite{Mohapatra:1986aw,Mohapatra:1986bd} and Linear~\cite{Akhmedov:1995ip,Malinsky:2005bi} seesaws, which allow for experimentally accessible  HNL mixing values. 

On the other hand, given their heavy scale, only the LHC or the proposed very high-energy colliders are capable of directly producing such HNLs. The LHC is actively searching for them (see, {\it e.g.},~Refs.~\cite{Abada:2022wvh,Marcano:2024bjs} for a summary), covering a mass range from a few GeV up to almost 1~TeV, above which the direct resonant s-channel production is highly suppressed. 
Interestingly, it has been proposed~\cite{Fuks:2020att} that HNL-mediated t-channel processes, in particular the lepton number violating (LNV) $W^\pm W^\pm \to \ell^\pm \ell^\pm$ ones, can extend this reach to higher masses. This search has been performed by ATLAS and CMS, deriving the first collider bounds for TeV HNLs~\cite{CMS:2022hvh,ATLAS:2023tkz,ATLAS:2024rzi}.

Nevertheless, t-channel processes must be handled with care, since, contrary to the resonant production, all heavy and light states can contribute coherently. As we will show in this work, in the context of the type-I seesaw model, the light neutrino contribution does indeed play a very important role and must be taken into account. 
In particular, we will show that it can cancel the HNL contribution in the LNV channel, suppressing the whole signal in a GIM-like fashion~\cite{Glashow:1970gm}, in the same way as for neutrinoless double beta decays~\cite{Blennow:2010th,Abada:2018qok} or for lepton colliders~\cite{Rizzo:1982kn,Belanger:1995nh,Rodejohann:2010jh}.

This suppression shifts our attention to the lepton number conserving (LNC) $W^+W^-\to\ell^+\ell^-$ process, where a very similar interplay between light and heavy neutrinos takes place to ensure again the correct unitary behavior. This fact has been recently pointed out in Refs.~\cite{Gabrielli:2026cjk,Chachava:2026mbu} for the lepton flavor conserving (LFC) channel and mainly for the inverse $\ell^+\ell^-\to W^+W^-$ process at lepton colliders, although the discussion is completely analogous. See also Ref.~\cite{Cvetic:2026pyt} for a recent similar discussion in $e\mu$ colliders. 

Our focus here will be studying this process at the LHC for three reasons. First, the $W^+W^-\to \ell^+_\alpha\ell^-_\beta$ process allows us to consider the lepton flavor violating (LFV) channel, whose cancellation is slightly simpler to study as it does not compete with other SM processes. Second, the LFV nature of the signal, together with the vector boson scattering (VBS) kinematics, generates a characteristic low-background signal: two energetic leptons of different flavor, with two VBS-like jets and low missing energy. Finally, our motivation is to show that this new search is able to improve collider limits already with current data. Nevertheless, we emphasize that our discussion can be easily generalized to future lepton and hadron high-energy colliders. 

The paper is organized as follows. In \cref{Sec:Theo} we briefly review the theoretical framework, introducing our notation and especially the two kinds of unitarity relations that will be relevant to our t-channel observables. In \cref{Sec:SvsT} we compare s- vs t-channel HNL searches, discussing their differences and advantages to probe different regions of the parameter space. Section~\ref{Sec:Cancellations} is devoted to study the t-channel in detail, with special emphasis on the importance of light neutrinos in both LNV and LNC processes. This section also motivates LNC over LNV t-channel searches, which we study in detail in \cref{Sec:EMU} simulating the $e^\pm\mu^\mp jj$ t-channel at the LHC. Our results prove that this search is promising to extend the LHC reach above the TeV scale, as concluded in \cref{Sec:concl}.

 %%%%%%%%%%%%%%%%%%%%%%%%%%%%%%%%%%%%%%%%%%%%%%%%%%%%%%%%%%%%%%%%%
 \section{Theoretical framework}
 \label{Sec:Theo}
 
The Lagrangian for a type-I seesaw model introducing $n$ right-handed neutrino fields is given by
\begin{equation}\label{eq:Lag}
\mathcal L_{\rm Type-I} = -Y_\nu \bar L \tilde\Phi \nu_R - \frac12 \overline{\nu_R^c} m_M \nu_R + {\rm h.c.}\,,
\end{equation}
which leads, after EW symmetry breaking, to a neutrino mass matrix 
\begin{equation}\label{eq:Mnu}
\mathcal M_\nu = \left(\begin{array}{cc} 0 & m_D \\ m_D^T & m_M\end{array}\right),
\end{equation}
in the $(\nu_L^c,\nu_R)^T$ basis and with $m_D=vY_\nu/\sqrt2$. The physical neutrinos are obtained diagonalizing this mass matrix via a unitary rotation $U$,
\begin{equation}
U^{T} \mathcal M_\nu U = \mathcal M_{\rm diag} = {\rm diag}(m_\nu,m_N)\,,
\end{equation}
generating light  $m_\nu$ and heavy $m_N$ neutrino masses. 
The complete rotation $U$ is usually decomposed in light and heavy blocks, 
\begin{equation}
U = \left(\begin{array}{cc} N & V \\ X & Y \end{array}\right).
\end{equation}
Here, $V$ is a $3\times n$ matrix encoding the mixings between the physical HNLs and the active neutrino flavors, which are their only connection to the SM particles and therefore the relevant phenomenological parameters together with their masses $m_N$. 
$N$ is the $3\times3$ mixing matrix between light neutrinos, thus taking the role of the PMNS matrix, with the important difference of not being unitary due to the presence of new neutrinos. In fact, it can be parametrized to make this property more explicit as~\cite{Fernandez-Martinez:2007iaa}
\begin{equation}
N=(\mathds 1-\eta)\,U_{\nu}\,,
\end{equation}
where $U_\nu$ is a $3\times3$ unitary matrix $(U_\nu\sim U_{\rm PMNS})$ and $\eta$ is a Hermitian matrix encoding the deviations from unitarity. Interestingly, this $\eta$ matrix can be mapped to the only dimension 6 effective operator that the type-I seesaw model generates when integrating the HNLs out at tree level~\cite{Broncano:2002rw}:
\begin{equation}\label{eq:dim6}
\eta = \frac12\,C_{d=6} = \frac12\, V V^\dagger\,, 
\end{equation}
which encodes precisely the low-energy effect that allows us to set constraints on the heavy HNL regime from flavor and EW precision observables~\cite{Blennow:2023mqx}.

The unitarity of the complete mixing matrix $U$ defines the first closure relation relevant for our processes, in particular for the LNC ones:
\begin{equation}\label{eq:uniLNC}
\sum_{i=1}^{3+n}U_{\alpha i}\, U^{*}_{\beta i} = \delta_{\alpha\beta}\,.
\end{equation}
On the other hand, the relevant relation for LNV processes will be
\begin{equation}\label{eq:uniLNV}
\sum_{i=1}^{3+n} U_{\alpha i} m_i U_{\beta i} = 0\,,
\end{equation}
where the sum runs again over all light and heavy neutrinos and $\alpha,\beta=e,\mu,\tau$. This relation, originating from the diagonalization of $\mathcal M_\nu$, is a direct implication of  $SU(2)_L$ invariance that imposes the zero entry in the mass matrix shown in \cref{eq:Mnu}. 
Therefore, both \cref{eq:uniLNC,eq:uniLNV} are completely general and apply to any kind of model with such a neutrino mass matrix. This includes, in particular, the low-scale seesaw models, developed to allow for phenomenologically accessible HNL mixings without spoiling light neutrino masses\footnote{Since in these realizations HNLs are introduced in pairs of opposite LN, the cancellation in \cref{eq:uniLNV} can be seen at first as a partial cancellation between the HNLs up to $\mathcal O(m_\nu)$, which is then canceled with the light neutrino contribution.}. 

For our numerical simulations, we will consider a simplified scenario with only one HNL, either Dirac or Majorana, as introduced in Ref.~\cite{Coloma:2020lgy}. In the case of a Majorana HNL, one of the light neutrinos will have a mass given by the seesaw relation, $m_{\nu_3}\sim V^2\, m_N$, while the other two will be massless. On the other hand, the Dirac HNL will effectively account for a low-scale seesaw scenario with one pair of heavy neutrinos forming a (pseudo-)Dirac pair, inducing only small light neutrino masses that we will neglect. While being too simple to account for neutrino oscillation data, these two scenarios already contain all the ingredients for a complete phenomenological collider analysis, including the relevant light-heavy neutrino interplay we are interested in.

 %%%%%%%%%%%%%%%%%%%%%%%%%%%%%%%%%%%%%%%%%%%%%%%%%%%%%%%%%%%%%%%%%
 \section{HNLs in s and t-channels at colliders}
 \label{Sec:SvsT} 
 
\begin{figure}[t!]
\centering
\includegraphics[width=0.49\textwidth]{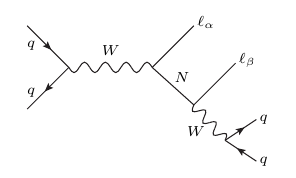}
\includegraphics[width=0.49\textwidth]{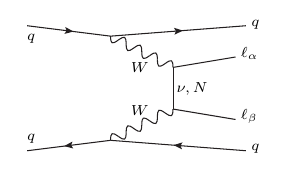}
\caption{HNLs in s- (left) and t- (right) diagrams at a proton-proton collider. Leptonic lines are shown without explicitly indicating their direction since these diagrams can be interpreted both in the LNC and in the LNV case. The t-channel also includes the light neutrino contribution, often neglected but relevant for this process, as discussed in the text. }\label{fig:SandTdiags}
\end{figure}

The phenomenology of heavy neutral leptons at high-energy colliders has been extensively studied over the past decades; for reviews, see Refs.~\cite{Deppisch:2015qwa,Cai:2017mow,Abdullahi:2022jlv}.
 Their production and decay occur via their mixings with active neutrino flavors $V_{\ell N} = 
 (V_{eN}, V_{\mu N}, V_{\tau N})$, which, in hadronic colliders, lead to diagrams such as those shown in \cref{fig:SandTdiags}, with the HNLs propagating in an s- or t-channel. In this section we briefly discuss the main characteristics  of these two kinds of processes. 
 
The s-channel shown in the diagram on the left of \cref{fig:SandTdiags} is the most commonly considered topology in hadron colliders. Its main and most interesting property is that it can directly produce the HNL, resonantly enhancing the total cross section. Using the narrow width approximation, we have
\begin{equation}
\sigma_s(pp\to \ell\ell j j) \simeq \sigma (p p \to \ell N)\, {\rm BR}(N\to \ell jj)\,,
\end{equation}
where we took as an example the semileptonic HNL decay leading to a dilepton signal, but the same applies for the trilepton one from the fully leptonic decay. The key point in any of these cases is that the resonant behavior reduces the scaling with the mixings in the cross section to $\mathcal O(V^2)$, instead of the naive $\mathcal O(V^4)$ obtained from counting vertices in \cref{fig:SandTdiags}. For this reason, the resonant s-channel production is the dominant one at hadron colliders as long as the HNL remains kinematically accessible, and has allowed the LHC to set the strongest constraints for GeV HNLs lighter than the $W$ boson~\cite{CMS:2024xdq}, surpassing LEP bounds~\cite{DELPHI:1996qcc} for $V_{eN}$ and $V_{\mu N}$ mixings. 
The resonant channel is also efficient in producing, via off-shell $W$ bosons\footnote{For HNLs of hundreds of GeVs the $W\gamma\to\ell N$ channel becomes relevant~\cite{Dev:2013wba,Alva:2014gxa}, although for the present discussion we just include it as part of the resonant s-channel production.}, HNLs below the TeV scale, above which they become too heavy for current collider energies. 

Another important property of the resonant production is that it introduces the total width of the HNL, $\Gamma_N$, as a key parameter. Besides reducing the powers of the mixing in the total cross section, it can also make the HNL long-lived enough to be probed through its displaced vertex signature, which enormously increases the LHC sensitivity in the $m_N\sim(1,20)$~GeV range~\cite{CMS:2022fut,ATLAS:2022atq,CMS:2023jqi}.
In contrast, the role of $\Gamma_N$ introduces a more involved dependence on the specific flavor pattern of the mixings, as it depends on all of the $V_{\ell N}$, making reinterpretations from commonly assumed single flavor dominance to a more realistic generic pattern quite challenging~\cite{Tastet:2021vwp,Abada:2022wvh}. 
This fact is even more relevant for displaced vertex searches, since the question of having a prompt or long-lived HNL may indeed depend on flavors apparently not involved in the process~\cite{Abada:2018sfh}.

As an alternative to the resonant s-channel, the HNLs can also mediate the t-channel process in the right diagram of \cref{fig:SandTdiags}. Being a non-resonant process, it is subdominant with respect to the s-channel when the HNLs are kinematically accessible, but it can lead to larger cross sections for heavier masses, as pointed out by Ref.~\cite{Fuks:2020att}. 
The reason is that, in general, t-channels suffer from a milder high-energy suppression than s-channels. Additionally, this particular process benefits from the high-energy enhancement of longitudinal $W$ boson scattering, thus providing a complementary test for masses above the TeV. 
Another interesting property is that it does not depend on $\Gamma_N$, avoiding all the difficulties mentioned above relating to the mixing flavor pattern.  

Nevertheless, the t-channel has two important caveats with respect to the resonant s-channel. First of all, it scales with the mixing\footnote{We note that this is the same scaling than the non-resonant s-channel. Nevertheless, the latter remains subdominant in the heavy HNL regime, since it does not benefit from the VBS kinematics.} as $\mathcal O(V^4)$, which means that it will be relevant to test relatively large mixings.  As we will see, the LHC could probe mixings of the order of $V^2\sim10^{-2}-10^{-1}$ via this kind of t-channels; however, this strong dependence\footnote{The lepton number and flavor conserving t-channel can actually reduce this scaling due to its interference with the SM contribution, see \cref{Sec:LNC}, although at the price of enhancing the backgrounds.} on the mixing makes it difficult to improve the bounds just by increasing luminosity. Thus, in order to explore smaller mixings, it seems more promising to make those heavy HNLs accessible by reaching higher energies.

Secondly, in the t-channel we must add contributions of all light and heavy neutrinos coherently, since there is no resonant behavior to tell them apart kinematically. As we will discuss in the next section, this is a crucial point in order to properly account for the correct behavior of the model. 

% %%%%%%%%%%%%%%%%%%%%%%%%%%%%%%%%%%%%%%%%%%%%%%%%%%%%%%%%%%%%%%%%%
 \section{The role of light neutrinos in t-channel processes}
 \label{Sec:Cancellations}

 In this section we discuss in detail the interplay between light and heavy type-I seesaw neutrinos in t-channel processes, which we separate in the lepton number violating and conserving cases due to their different kind of cancellations.
 For concreteness, we focus on the $WW\to \ell\ell$ scattering, the relevant one for the LHC, although our conclusions are easily  extended to other kinds of colliders. 
 Moreover, we present our discussion at the hard process level --since it already contains all the relevant ingredients-- which is then translated to a hadronic collider by means of VBS configurations.

% %%%%%%%%%%%%%%%%%%%%%%%%%%%%%%%%%%%%%%%%%%%%%%%%%%%%%%%%%%%%%%%%%
 \subsection{Lepton Number Violating t-channels}
 \label{Sec:LNV}

Lepton Number violation is the smoking-gun signature of Majorana HNLs~\cite{Atre:2009rg} and, as such, the $W^\pm W^\pm\to\ell^\pm\ell^\pm$ process has been studied~\cite{Fuks:2020att} and searched for~\cite{CMS:2022hvh,ATLAS:2023tkz,ATLAS:2024rzi} at the LHC, leading to bounds of $V^2\lesssim \mathcal O(10^{-1})$ for $m_N\sim1$~TeV. 
The crossed process $\ell^-\ell^-\to W^-W^-$, often referred to as inverse neutrinoless double-beta decay, has also been studied at future lepton colliders~\cite{Rizzo:1982kn,Belanger:1995nh,Rodejohann:2010jh,Wang:2016eln,Jiang:2023mte,Li:2023lkl,deLima:2024ohf}.

\begin{figure}[t!]
\centering
\includegraphics[width=0.49\textwidth]{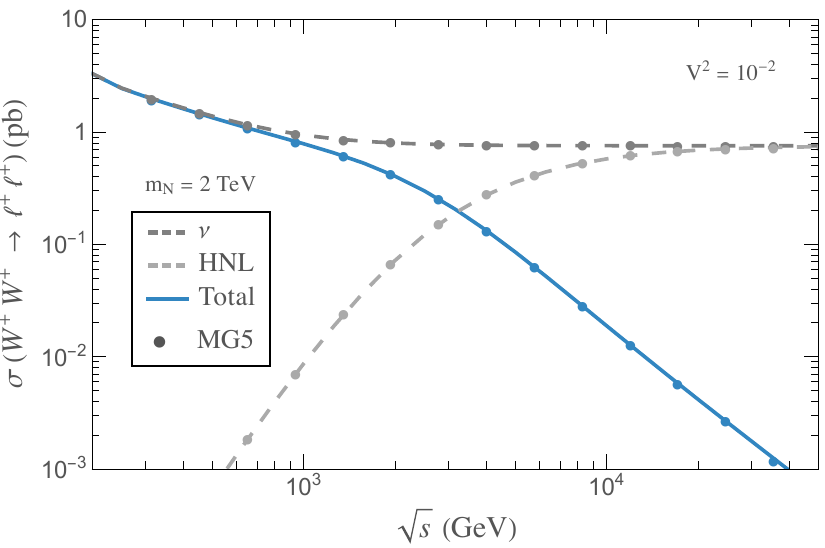}
\includegraphics[width=0.49\textwidth]{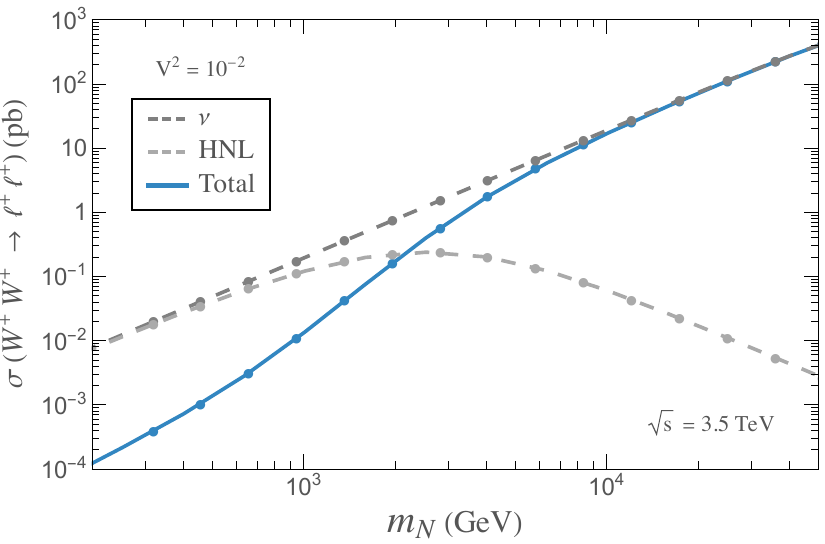}
\caption{LNV $W^+W^+\to \ell^+\ell^+$ process in a simplified scenario with one light neutrino $\nu$ and one HNL, whose contributions are displayed separately (dashed lines) to show their cancellation when computing the total contribution (solid line). A cut on $|\eta_\ell| < 5$ has been imposed to avoid numerical instabilities in the integration.  Lines are obtained analytically, while dots show how our treatment with {\tt MadGraph5} is able to reproduce them.}\label{fig:LNVsubprocess}
\end{figure}

Some of these works consider only the HNL mediation in the right diagram of \cref{fig:SandTdiags}, neglecting the light neutrino contribution. A priori, this seems to be a reasonable assumption, since LNV processes are proportional to the mass of the Majorana neutrino, which should be tiny for active neutrinos. The problem is that a single Majorana HNL with a mass and mixing as those probed by these analyses generates a large mass for light neutrinos, 
\begin{equation}\label{eq:mnu100}
    m_\nu \sim V^2\, m_N \sim 100~{\rm GeV}\,,
\end{equation}
and such a not-so-light neutrino actually dominates the whole LNV processes, since it is less suppressed that the HNL contribution. 
This is shown numerically in \cref{fig:LNVsubprocess} for a simplified model with only one HNL $N$ and one light neutrino $\nu$ with mass as predicted by the model.

The type-I seesaw actually predicts much more than just a large active neutrino contribution, since it also leads to a cancellation between the heavy and light contributions. In order to show this generic property, we consider the generic type-I seesaw model with $n$ HNLs. Within this setup, the t-channel contribution for the LNV $W^+W^+\to \ell^+_\alpha\ell^+_\beta$ process is given by 
\begin{equation}
    \mathcal M_t = \sum_{i=1}^{3+n} U_{\alpha i}\,U_{\beta i}\,\frac{m_i}{t-m_i^2}\,\mathcal A_t(s,\cos\theta)\,,
\end{equation}
where $\mathcal A_t$ contains all the terms independent of the neutrino parameters. The u-channel contribution, $\mathcal M_u$, is equivalent, with a total total amplitude of $\mathcal M_{\rm LNV}= \mathcal M_t + \mathcal M_u$. From this factorization, we can clearly distinguish two limits for the HNL contribution, which explain the overall behavior of the total LNV rate in \cref{fig:LNVsubprocess}: 
\begin{itemize}
    \item When all HNLs are heavier than the collider energy ($m_N^2\gg s$) their contribution is suppressed, leaving only the active neutrino one. This is nothing but integrating out the HNLs and generating the Weinberg operator~\cite{Weinberg:1979sa} at low-energies, which has been already searched for\footnote{Note that the bounds derived by these collider tests of the Weinberg operator actually exclude the $m_\nu$ values in \cref{eq:mnu100}, already indicating that the whole process is dominated by the active neutrino contribution.} at the LHC~\cite{CMS:2022hvh,ATLAS:2023tkz,ATLAS:2024rzi} as originally proposed by Ref.~\cite{Fuks:2020zbm}. We note that the growth of this contribution in the right panel of \cref{fig:LNVsubprocess} is due to the seesaw relation $m_\nu\sim V^2 m_N$ with fixed mixing, although it vanishes in the $m_N\to \infty$ limit (since $m_\nu\to\infty)$, not shown within the range of the figure.

    \item When all HNLs are lighter than the collider energy ($m_N^2\ll s$) the amplitude becomes proportional to 
    \begin{equation}
        \mathcal M_{\rm LNV}\sim \sum_{i=1}^{3+n} U_{\alpha i}\,U_{\beta i}\,m_i = 0\,,
    \end{equation}
    which vanishes due to the seesaw relation in \cref{eq:uniLNV}, strongly suppressing LNV rates. 
 
\end{itemize}

Consequently, the LNV t-channel is not the best process to learn about Majorana HNLs from a type-I seesaw model, since their role is either to suppress the total rate or to give a subdominant contribution. The only potentially relevant HNL contribution could come from a model with several of them, some above and some below the characteristic energy of the process, analogously to their phenomenology for neutrinoless double beta decays~\cite{Blennow:2010th,Abada:2018qok}. Nevertheless, such a scenario requires a non-minimal model that goes beyond the scope of our current analysis. 

\vspace{.5cm}
To conclude this section, we emphasize that the origin of such sizable active neutrino contribution is the large $m_\nu$ generated by the single Majorana HNL, which is clearly excluded by neutrino data. As already mentioned, low-scale seesaw realizations provide a viable framework for large mixings without inducing large values for $m_\nu$, but, unfortunately, they require an approximate lepton number symmetry~\cite{Moffat:2017feq} which strongly suppresses the LNV signatures. 
It is true that the latter could be avoided if the low-scale HNLs had the correct values as to oscillate between them before decaying~\cite{Antusch:2017ebe,Drewes:2019byd,Fernandez-Martinez:2022gsu}, but this would require resonantly producing them and therefore does not apply to the t-channel. 

It is still common practice to perform HNL collider analyses using a phenomenological model considering $m_N$ and $V_{\ell N}$ as completely independent parameters, assuming that there will be some kind of mechanism suppressing the potentially large contributions to $m_\nu$ without modifying the phenomenology of the HNL itself.
While this approach is partially justified for resonant s-channel searches, it is less obvious for the t-channel, where all contributions must be added coherently and, as we have just seen, already a type-I seesaw mechanism, which is the original motivation for the existence of the HNLs, leads to a strong suppression of the LNV channels. 
%Avoiding this supression would require extending the model to generate a Majorana mass for the $\nu_L$ fields, such as in the type-II seesaw model, but this would in general modify also the HNL phenomenology.
For all these reasons, we conclude that the LNV t-channel process is not a promising signal to probe the type-I seesaw mechanism, neither at a hadronic nor at a leptonic collider, and from now on we will focus instead in the LNC processes. 

As a final technical remark we note that, in order to reproduce correctly the light-heavy interplay during event simulation, it is crucial to use a model that treats the unitarity of the whole neutrino mixing matrix properly, as recently pointed out by Ref.~\cite{Chachava:2026mbu}. In the case of the LNV process, moreover, it is also important to introduce the seesaw prediction for $m_\nu$.
For this reason, the analytical results in \cref{fig:LNVsubprocess} were reproduced with {\tt MadGraph5} using the UFO model provided by Ref.~\cite{Coloma:2020lgy} with a small modification to include $m_\nu$.

% %%%%%%%%%%%%%%%%%%%%%%%%%%%%%%%%%%%%%%%%%%%%%%%%%%%%%%%%%%%%%%%%%
 \subsection{Lepton Number Conserving t-channels}
 \label{Sec:LNC}

 LNC processes are generated by both Dirac and Majorana HNLs. From the experimental side, they could be less appealing due to higher SM backgrounds, but they are very relevant from the theory side, since they could probe the large mixings allowed by the low-scale seesaw models with suppressed LNV signals. This is also true for t-channel mediated processes.

 The neutrino contribution to the LNC $W^+W^-\to\ell^+_\alpha\ell_\beta^-$ process takes place, once more, through the diagram on the right of \cref{fig:SandTdiags}, but now the dependence of the amplitude on the neutrino parameters is different,
 \begin{equation}\label{eq:LNCfullM}
     \mathcal M_{\rm LNC}^n = \sum_{i=1}^{3+n} U_{\alpha i} U^{*}_{\beta i} \frac{\slashed q}{t-m_i^2}\, \mathcal B(s,\cos\theta)\,, 
 \end{equation}
 where $q$ is the momentum of the neutrino propagator, $q^2=t$, and $\mathcal B$ contains the rest of the amplitude independent of the neutrino parameters. 
 We note that the LNC amplitude is not proportional to the neutrino mass, implying that we must always take into account both light and heavy neutrinos, independently of how  we are treating the active neutrino mass generation. In fact, from now on we will take $m_\nu\simeq0$, which can be justified as considering Dirac HNLs, also leading to LNC signatures.
 
 As before, we can distinguish two limits,
\begin{itemize}
    \item For heavy HNLs ($m_N^2\gg s$), the only relevant contribution is that of the active neutrinos, which enters in the form of the non-unitarity matrix $\eta$:
    \begin{equation}\label{eq:LNCheavyN}
    \mathcal M^{\nu}_{\rm LNC}\simeq \sum_{i=1}^{3} U_{\alpha i} U^{*}_{\beta i} \, \mathcal M_{\rm SM}^{\nu}= 2\eta_{\alpha\beta}\,\mathcal M_{\rm SM}^{\nu}\,,
    \end{equation}
    where $\mathcal M_{\rm SM}^{\nu}$ stands for the amplitude for massless SM neutrinos. 
    This is again the low-energy effect of integrating the HNLs out, but in this case leading to the dimension 6 operator. In fact, at high-energies $\mathcal M_{\rm SM}^{\nu}\propto s$, so the LNC amplitude has the $s$ dependence expected from a dim-6 operator.  

    \item For lighter HNLs ($m_N^2\ll s$), both light and heavy HNLs contribute equally, recovering the SM prediction due to the unitarity of the mixing matrix, 
    \begin{equation}\label{eq:LNClightN}
        \mathcal M_{\rm LNC}^n\simeq \sum_{i=1}^{3+n} U_{\alpha i}\,U^{*}_{\beta i} \,\mathcal M_{\rm SM}^{\nu}= \delta_{\alpha\beta}\,\mathcal M_{\rm SM}^{\nu}\,.
    \end{equation}
    Note that it also grows linearly with $s$ at high energies.
\end{itemize}

In order to discuss the implications of the above equations, we must distinguish between the lepton flavor conserving and violating channels. 
For the former, the SM also contributes to the process via s-channel photon and $Z$ exchange\footnote{Also a subdominant contribution from Higgs boson exchange.}, which compete with the neutrino contribution.
In the SM, both $\mathcal M_{\rm SM}^{\nu}$ and $\mathcal M_{\rm SM}^{\gamma/Z}$ amplitudes grow with $s$, but their potentially dangerous contributions cancel each other out, leading to the correct high-energy behavior. This is still true in the case of light HNLs in \cref{eq:LNClightN}, but not for heavy HNLs introducing a non-unitarity effect only for the neutrino contribution as in \cref{eq:LNCheavyN}, since they spoil the exact cancellation with the $\gamma/Z$ diagrams and therefore generate a bad high-energy behavior. 

This pathological growth with energy due to the non-unitarity of the light neutrino mixing matrix, potentially leading to violations of perturbative unitarity, has been recently pointed out by Refs.~\cite{Gabrielli:2026cjk,Chachava:2026mbu}. These works mostly focus on the inverse $\ell^+\ell^-\to W^+W^-$ scattering at lepton colliders, although the discussion is completely analogous for the $W^+W^-\to \ell^+\ell^-$ case at hadron colliders, as discussed in Ref.~\cite{Gabrielli:2026cjk}. The main idea is to measure this scattering process at different energies with the goal of observing a growth in the cross sections deviating from the SM behavior at high energies. This has the interesting advantage of being sensitive also to an interference term with the SM, which reduces the dependence on the HNL mixing to $\mathcal O(V^2)$ instead of the $\mathcal O(V^4)$ of the t-channel (or equivalently $\mathcal O(\eta)$ instead of $\mathcal O(\eta^2)$), and therefore it is more promising to probe smaller mixings. Its main drawback is that it obviously has large SM backgrounds, which we actually need to measure precisely and at different energies.

The discussion becomes simpler for the LFV channels, since there is no SM contribution to compete with and therefore the whole amplitude is given by \cref{eq:LNCfullM}. 
As an illustration, let us compute the scattering of longitudinal gauge bosons, the dominant contribution at energies well above the $W$ boson mass. 
Taking all the HNLs to be degenerate, we can further simplify it using \cref{eq:uniLNC}
\begin{equation}
    \sum_{i=4}^{3+n} U_{\alpha i} U^{*}_{\beta i} = -\sum_{i=1}^{3} U_{\alpha i} U^{*}_{\beta i} = -2\eta_{\alpha\beta}\,,
\end{equation}
and express the cross section in terms of the off-diagonal entry of the non-unitary matrix $\eta$.
Then, neglecting the $W$ boson mass, we have
\begin{equation}\label{sigmaLLnoMw}
\sigma(W_L^+W_L^-\to \ell^+_\alpha\ell^-_\beta)\simeq~\frac{g^4}{144\pi}\frac{m_N^4}{m_W^4}|\eta_{\alpha\beta}|^2\frac{1}{s}\left\{ \Bigg(1+\frac{2m_N^2}{s}\Bigg)\log{\left[ \frac{s+m_N^2}{m_N^2} \right]}-2 \right\}\,. 
\end{equation}
It is also interesting to analyze the light and heavy HNL limits, 
\begin{align}
\sigma (m_N^2\ll s) &\approx~\frac{g^4}{144\pi}\frac{m_N^4}{m_W^4}|\eta_{\alpha\beta}|^2\, \frac{1}{s}\log\frac{s}{m_N^2}\, ,\label{mNSmall}\\
\sigma (m_N^2\gg s)&\approx~\frac{g^4}{864\pi\, m_W^4}|\eta_{\alpha\beta}|^2\, s\, .\label{mNLarge}
\end{align}
In the heavy HNL limit, they are effectively integrated out, so the cross section grows with $s$ following the non-unitarity dim-6 operator for active neutrinos. When the energy is large enough, the HNLs also become relevant, restoring the unitarity of the neutrino mixing matrix and curing the pathological energy behavior. 

\begin{figure}[t!]
\centering
    \includegraphics[width=0.49\textwidth]{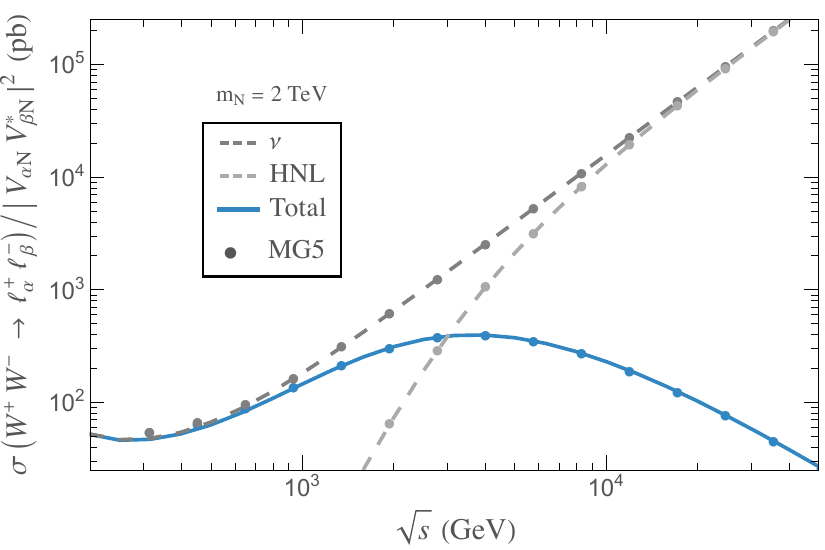}
    \includegraphics[width=0.49\textwidth]{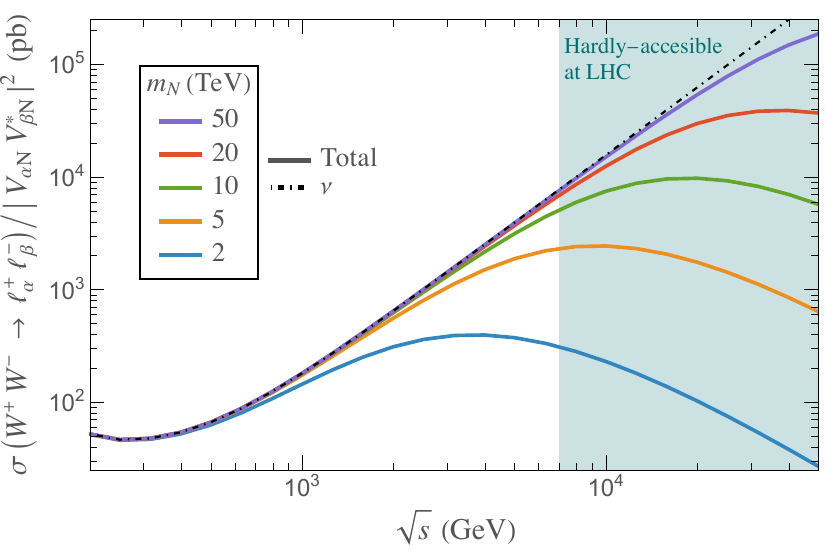}
\caption{LFV $W^+W^-\to \ell^+_\alpha\ell^-_\beta$ process in a type-I seesaw with degenerate HNLs. Left panel is analogous to that of \cref{fig:LNVsubprocess}, with the same notation. The right panel shows the total cross section for different values of the HNL masses. The shadowed blue region indicates the high-energy region outside the LHC reach, only for illustration purposes, roughly estimated to (subprocess) energies of 7~TeV.}\label{fig:LFVsubprocess}
\end{figure}

This interplay between the light and heavy neutrinos can be seen in \cref{fig:LFVsubprocess}, now for the complete $W^+W^-\to \ell^+_\alpha\ell^-_\beta$ scattering. From the left panel, it is again manifest the importance of considering all neutrinos to recover the correct behavior with energy. In the right panel we see how this competition happens for different masses. The cross section follows the light neutrino contribution at lower energies (the dim-6 operator), until the HNL becomes relevant and suppresses the ratio. The heavier the HNL, the later it enters to restore the unitarity, so the light neutrino contribution has more space to grow and the total cross-section becomes larger. In practice, however, this counterintuitive behavior saturates when the HNL becomes too heavy for the energy of the collider, which in this figure we took to 7~TeV as a crude estimate of the maximal energy for a VBS scattering at the LHC. For such heavy HNLs, the collider only sees the light neutrino contribution, independently of the heavy mass, probing  the low-energy dim-6 operator.

This figure also shows that the LFV t-channel contribution is promising to observe the effects of very heavy HNLs at the LHC, leading to high-energy dileptons with opposite flavor together with jets fulfilling the characteristic VBS kinematics. 
We will show the potential of this channel in the next section. 

% %%%%%%%%%%%%%%%%%%%%%%%%%%%%%%%%%%%%%%%%%%%%%%%%%%%%%%%%%%%%%%%%%
 \section{Case of study: high-energy $\boldsymbol{e\mu jj}$ t-channel at the LHC}
 \label{Sec:EMU}

 In this section we study in detail the potential of the LHC to search for the LFV t-channel induced by type-I seesaw heavy neutrinos, showing that this is a viable process at colliders when the HNLs are very heavy. For concreteness, we will focus only on the $e-\mu$ channel, although the same idea translates to the $\tau-\ell$ channels. 

 The HNLs leading to LFV t-channels can be either Dirac or Majorana, both leading to the same result as long as $m_\nu$ remains negligible. This means that this search is sensitive to the low-scale seesaw mechanisms justifying the large mixings the LHC will be sensitive to, since these models predict pseudo-Dirac HNLs. 
 In order to simplify our simulation setup, we will consider the existence of only one HNL, although our results actually apply to an arbitrary number of HNLs as long as they are degenerate in mass and that we replace the individual mixings by the sum of all mixings, as 
 \begin{equation}
     \big|V^{}_{\alpha N}V^*_{\beta N}\big| \longrightarrow \Big|\sum_{i=1}^{n} V^{}_{\alpha N_i} V^*_{\beta N_i}\Big|\,.
 \end{equation}
The general idea is the same for non-degenerate HNLs and can therefore be easily generalized, keeping in mind that the complete restoration of unitarity will not happen unless all the HNLs fall below the energy of the process. 

 The signal we are searching for corresponds to the $WW$ scattering in the right panel of \cref{fig:SandTdiags}, which, at hadronic colliders, can be tagged by the characteristic kinematics of the two outgoing jets after emitting two $W$ bosons. With this in mind, we define our basic selection criteria as containing exactly one electron or positron, exactly one muon or antimuon and at least two jets. 
 %We additionally veto events with $b$-jets, as to reduce backgrounds from tops. 

 The main backgrounds will come from $t\bar t$ and $VVjj$ production, with $V=W,Z$ bosons. The former is a dominant background due to its large cross section at the LHC, but it can be strongly suppressed requiring a VBS configuration of the final jets and vetoing events with $b$ jets. The $VVjj$ process directly contributes to VBS kinematics, but the corresponding final states, obtained via leptonic decays of the vector bosons, contain neutrinos --or higher multiplicities of charged leptons--. Therefore, requiring a low missing transverse energy will be another important discriminant in our analysis. 

 Other potential backgrounds could arise from multijet and $W$+jet processes due to the misidentification of jets as leptons. Nevertheless, these are usually estimated from data and we will not include them in our analysis. 
 Continuous dilepton and trilepton processes, $2\ell$+X and $3\ell\nu+X$ could also mimic our LFV signal, but we have estimated them to be subdominant in the high-energy region ($m_{\ell\ell}>200$~GeV) in which we will be interested.

 We perform our simulation using {\tt MadGraph5}~\cite{Alwall:2014hca,Frederix:2018nkq} to generate parton level events, {\tt Phythia8}~\cite{Bierlich:2022pfr} for the showering, {\tt FastJet}~\cite{Cacciari:2005hq,Cacciari:2011ma} for clustering and {\tt Delphes3}~\cite{deFavereau:2013fsa} for detector simulation.
 The signal is generated at leading order (LO) using the  FeynRules~\cite{Christensen:2008py,Alloul:2013bka} UFO~\cite{Degrande:2011ua} for the type-I seesaw as implemented by Ref.~\cite{Coloma:2020lgy}, which introduces the full neutrino unitary mixing matrix\footnote{In contrast to the  {\tt HeavyN}~\cite{Pascoli:2018heg} implementation, commonly used in resonant searches, as pointed out recently by Ref.~\cite{Chachava:2026mbu}.} and therefore reproduces the correct energy behavior discussed in \cref{Sec:LNC}. We only consider the t-channel in the generation, since the s-channel only leads to a subdominant contribution to VBS configurations, even for light masses.
 
The $t\bar t$ background is generated at next-to-leading order (NLO) in QCD and up to one additional jet matching following the {\it FxFx} procedure~\cite{Frederix:2012ps}.
Moreover, given its large cross section of around 1~nb and that we will be interested in high-energy events, we optimize the generation of events doing it in non-overlapping windows of $m_{t\bar t}$, as explained for instance in the appendix of Ref.~\cite{Arganda:2019gnv}. The top quarks are then decayed with {\tt MadSpin}~\cite{Artoisenet:2012st} following the $t\to Wb, W\to \ell\nu$ channel. 

The simulation of the $VVjj$ background is actually separated in two, following Ref.~\cite{Fuks:2020att}: a mixed EW-QCD contribution and a purely EW one. The former gives a larger contribution of $\mathcal O(\alpha^2 \alpha_s^2)$ to the total cross section, but it does not contain the VBS topologies, which are purely EW, {\it i.e.}, $\mathcal O(\alpha^4)$.  Therefore, we treat and generate these two contributions as two separated backgrounds, ensuring a better characterization of the smaller, pure EW contribution that will be more similar to the signal. 
In both cases, we generate them at LO, in windows of $m_{VV}$ and decaying the gauge bosons with {\tt MadSpin}. 

\begin{figure}[t!]
\centering
    \includegraphics[width=0.49\textwidth]{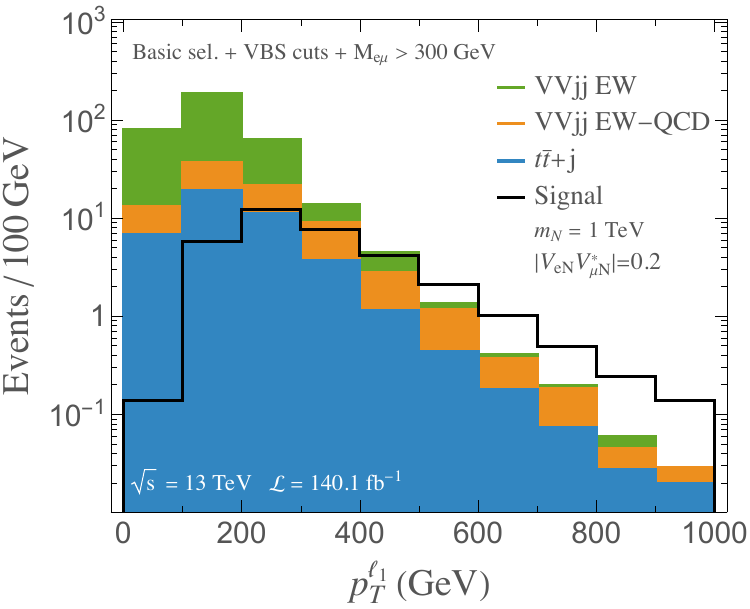}
    \includegraphics[width=0.49\textwidth]{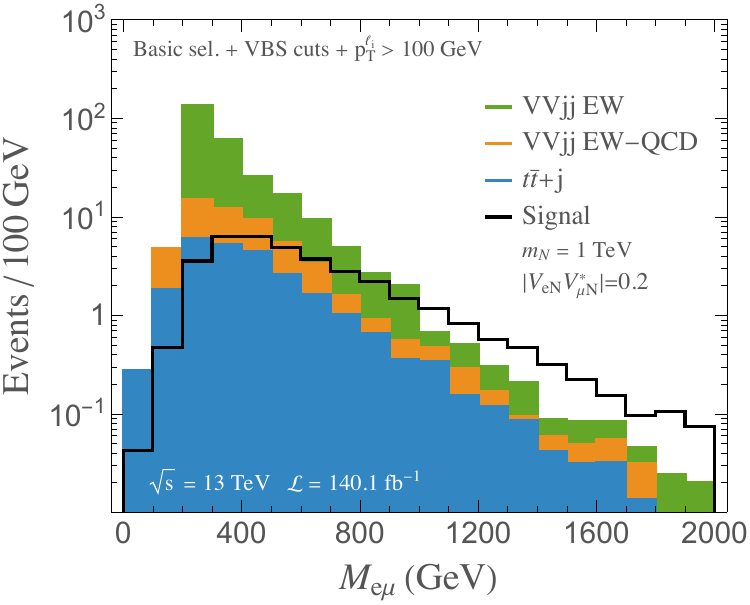}
\caption{Distributions of a representative signal benchmark and cumulative background events as a function of the transverse momentum of the leading lepton $p_T^{\ell_1}$ (left) and as a function of the dilepton invariant mass $M_{e\mu}$ (right). Basic selection and VBS cuts as in \cref{eq:BasicCut,eq:VBScuts} are implemented. Preliminary $M_{e\mu}$ and $p_T^{\ell_1}$ cuts are required, respectively, for a first background reduction, although we note these cuts are softer than in the final signal region.
Signal events populate the high energy regions more efficiently than the backgrounds, pointing towards rather hard $p_T^{\ell_1}$ and $M_{e\mu}$ selection cuts.}\label{fig:Distributions}
\end{figure}

All in all, the signal we are interested in consists of $e\mu$ events and at least two additional jets, with the two leading jets fulfilling the VBS kinematics cuts. More precisely, we define our basic selection and VBS cuts as:
\begin{align}
    \text{Basic sel.}:&\quad N_{e^\pm}=1\,,\quad N_{\mu^\pm} = 1\,,\quad N_j\geq2 \,,\quad N_b=0\,,\label{eq:BasicCut}\\
    \text{VBS cuts}:&\quad p_T^{j_1}>40~{\rm GeV}\,,\quad p_T^{j_2}>20~{\rm GeV}\,,\quad |\Delta\eta_{j_1j_2}|>4\,,\quad M_{j_1j_2}> 600~{\rm GeV}\,,\label{eq:VBScuts}
\end{align}
where $j_{1(2)}$ refers to the (sub)leading $p_T$ jet.

\begin{figure}[t!]
\centering
    \includegraphics[width=0.95
\textwidth]{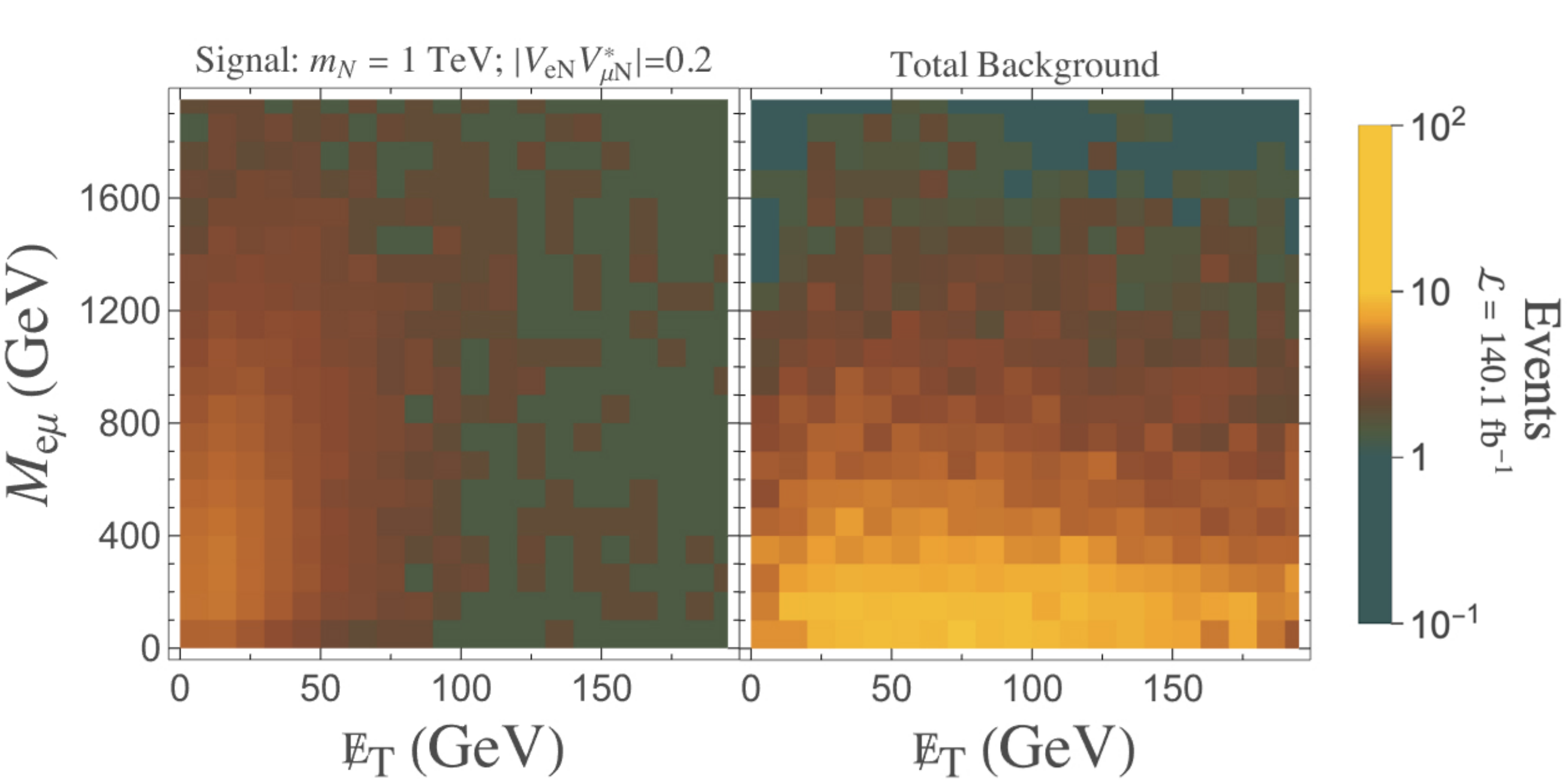}
\caption{Distributions of a representative signal benchmark and total background events as a function of the dilepton invariant mass $M_{e\mu}$ and the missing transverse energy $\slashed{E}_T$. Basic selection and VBS cuts as in \cref{eq:BasicCut,eq:VBScuts} are implemented. \textit{Left:} the signal events, not containing neutrinos in the final state, generally showcase small $\slashed{E}_T$, with $M_{e\mu}$ extending to rather large values. \textit{Right:} On the contrary, background events have a broad distribution in $\slashed{E}_T$ and, additionally, they tend to cluster in the lower $M_{e\mu}$ region. }\label{fig:Distributions2D}
\end{figure}

Moreover, as we saw in the previous section, the t-channel process leads to high-energy leptons with no associated missing energy. This is clearly visible in \cref{fig:Distributions,fig:Distributions2D}, highlighting the different features of the signal with respect to the backgrounds. Consequently, we define our {\bf signal region} as
\begin{equation}\label{eq:cutsSR}
    p_T^{\ell_1} > 300~{\rm GeV}\,,\qquad p_T^{\ell_2}> 250~{\rm GeV}\,, \qquad
    M_{e\mu}>600~{\rm GeV}\,,\qquad \slashed{E}_T<30~\rm{GeV}\,,
\end{equation}
where, again, $\ell_{1(2)}$ refers to the (sub)leading lepton.

In order to quantify the exclusion potential of this search, we follow Ref.~\cite{ATLAS:2020yaz} and compute the significance as
\begin{equation}
Z=\frac{(n-n_{B})}{|n-n_{B}|}\sqrt{2\left[ n\log{x}-(n_B^2/\delta_B^2)\log{y} \right]}\,,
\end{equation}
with
\begin{equation}
x=n(n_B+\delta_B^2)/(n_B^2+\delta_B^2)\,,\qquad
y=1+\frac{\delta_B^2}{n_B}(n-n_B)/(n_B+\delta_B^2)\,,
\end{equation}
and where $n=n_S+n_B$ is the total number of observed events within our signal region, being $n_S$ and $n_B$ the number of signal and total background events in said signal region at a given luminosity, respectively. The parameter $\delta_B$ accounts for the uncertainty in $n_B$, which we take, based on experimental measurements of $W^+W^-$ scattering processes \cite{CMS:2022woe}, at a value of $\delta_B=0.2$.

\begin{figure}[t!]
\centering
    \includegraphics[width=0.9\textwidth]{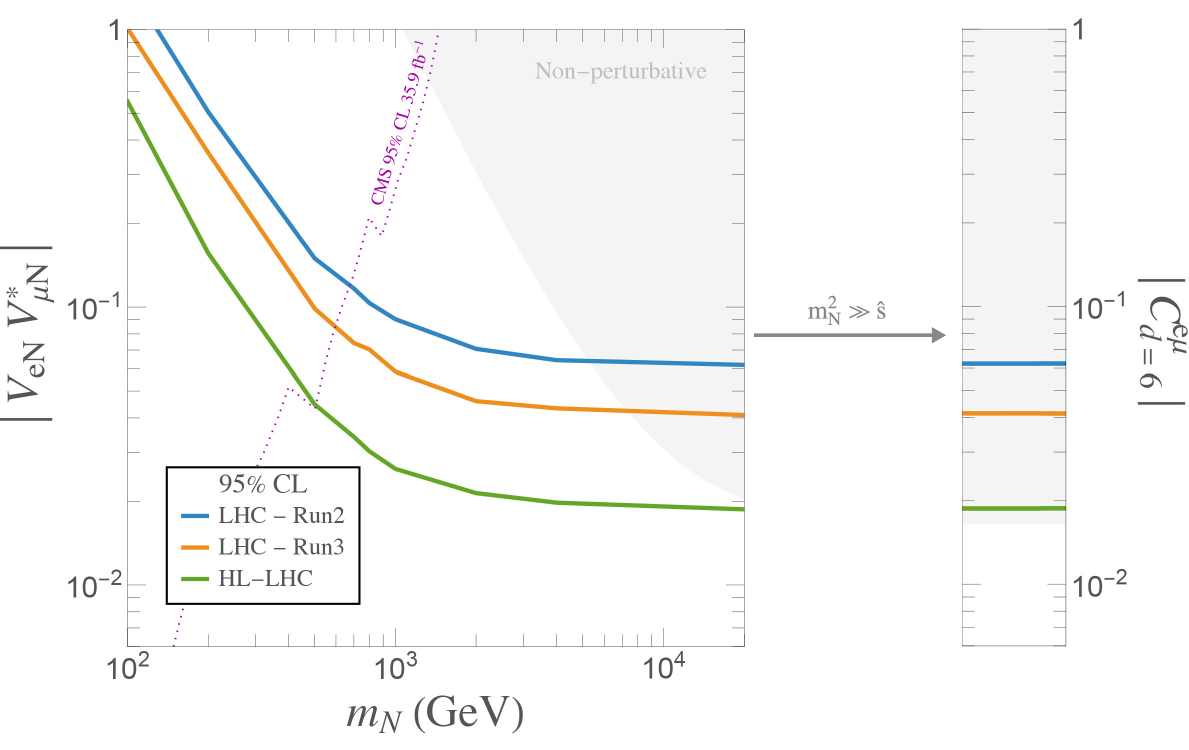}
\caption{95\% CL exclusion limits from the neutrino mediated LFV $e^\pm\mu^\mp jj$ t-channel at the LHC Run 2 (blue), Run 3 (yellow) and HL-LHC (green). The dotted line shows the CMS bound from s-channel same-sign dilepton ($e\mu$) search~\cite{CMS:2018jxx}, assuming $|V_{eN}|=|V_{\mu N}|$. In the gray area the $W^+W^-\to e^+\mu^-$ process violates perturbative unitarity. We also show how this search would be sensitive to the dimension 6 effective operator in \cref{eq:dim6}, obtained in the type-I seesaw after integrating the HNLs out.}\label{fig:MoneyPlot}
\end{figure}

The 95\% CL exclusions are shown in \cref{fig:MoneyPlot}, computed for the LHC Run 2 ($\sqrt s =13$~TeV, $\mathcal L=140.1~{\rm fb}^{-1}$~\cite{ATLAS:2022hro}) and Run 3 ($\sqrt s =13.6$~TeV, $\mathcal L=332~{\rm fb}^{-1}$~\cite{ATLAS:LumiRun3}) as well as for the future HL-LHC ($\sqrt s =14$~TeV, $\mathcal L=3000~{\rm fb}^{-1}$~\cite{HL-LHC}). For reference, we show the exclusion limits from the CMS search~\cite{CMS:2018jxx} for the resonant s-channel leading to the same-sign $e^\pm\mu^\pm$ signature, where we have assumed the flavor pattern $|V_{eN}|=|V_{\mu N}|$ (and $V_{\tau N}=0$) for concreteness. This search also explored the $e^\pm e^\pm$ and $\mu^\pm\mu^\pm$ channels, which set slightly weaker bounds at these masses after the proper recasting to the same flavor pattern~\cite{Abada:2022wvh}. We note that there are also more recent resonant searchers for heavy HNLs in the trilepton channels~\cite{CMS:2024xdq}, but their recasting from the single flavor hypothesis is more involved and moreover we expect them to provide substantially similar bounds. Lastly, let us stress again that the t-channel bounds are free from this flavor pattern dependency, as they directly depend on the $V^{}_{eN} V^*_{\mu N}$ combination. 

We also highlight in the figure the region where the validity of our perturbative treatment of the process could be violated. Indeed, since $V\sim Y_\nu v/m_N$, we are testing rather large values of the Yukawa coupling in Eq.~\eqref{eq:Lag}, for which our computation at leading order in perturbation theory\footnote{Moreover, in this large coupling region the total HNL width $\Gamma_N\propto V^2 m_N^3$ becomes comparable to its mass.} may break down~\cite{Urquia-Calderon:2024rzc}. In order to determine the parameter space compatible with perturbativity, we perform a partial wave decomposition of the matrix element of the subprocess $W^+_LW^-_L\to \ell^+_{\alpha}\ell^-_{\beta}$ and require it remains unitary within the energies we are testing, {\it i.e.},  up to $\sqrt{s}<14$~TeV. 
We note that the change of behavior of the non-perturbative region at high $m_N$ is due to the fact that we took into account the whole HNL mass dependence of the process, whose effect becomes visible for masses comparable to the cut-off energy. We recover the result of Ref.~\cite{Urquia-Calderon:2024rzc} when neglecting the $m_N$ effects, or, equivalently, when requiring that the process remains perturbative up to arbitrarily high energies. 

%\CGG{...this explains the non-linear behaviour...recuperar los bounds de Kevin}  As it can be seen in the plot, even though our sensitivity results extend above $10\rm \, TeV$, the required values for the mixing only render our predictions consistent up to $\sim 10\,\rm TeV$.

%becomes non-perturbative...\XM{tengo que coser este parrafo aqui}
%To close this section, we remark that in the parameter space we are probing, with large mixings and quite massive HNLs, the validity of our perturbative treatment of the process must be carefully addressed.  Indeed, since $V\sim Y_\nu v/m_N$, we are testing rather large values of the Yukawa coupling in Eq.~\eqref{eq:Lag}, for which our predictions of the process, computed at leading order in perturbation theory, may break down~\cite{Urquia-Calderon:2024rzc}. In order to determine the parameter space compatible with perturbativity, we perform a partial wave decomposition of the matrix element of the process $W^+_LW^-_L\to \ell^+_{\alpha}\ell^-_{\beta}$ taking into account the HNL mass dependence. \CGG{...this explains the non-linear behaviour...recuperar los bounds de Kevin} The results are shown as a shaded gray region in fig.~\ref{fig:MoneyPlot}. As it can be seen in the plot, even though our sensitivity results extend above $10\rm \, TeV$, the required values for the mixing only render our predictions consistent up to $\sim 10\,\rm TeV$.

From \cref{fig:MoneyPlot} it is clear that the LFV t-channel is a promising window to extend the LHC reach to higher masses, with better sensitivities than the resonant s-channel for $m_N\gtrsim800-1000$~GeV. Moreover, this signature does not rely on the Majorana nature of neutrinos, so it does probe also the existence of low-scale seesaw Dirac HNLs. The general shape follows that discussed in \cref{Sec:LNC}, with lower sensitivity at lower masses due to the cancellations in \cref{eq:uniLNC}, and saturating to a constant value at heavy masses when the HNLs are effectively integrated out. 
In other words, the horizontal asymptotes correspond to the LHC sensitivity to the dim-6 operator in \cref{eq:dim6}, as shown in the right panel of \cref{fig:MoneyPlot}, although unfortunately they lie inside the non-perturbative area for the energies and luminosities considered. All in all, we conclude that the LFV t-channel is the most promising signature to explore the $e\mu$ sector at the LHC for HNL masses between 1 and 10~TeV.

 %%%%%%%%%%%%%%%%%%%%%%%%%%%%%%%%%%%%%%%%%%%%%%%%%%%%%%%%%%%%%%
 \section{Summary and Conclusions}
 \label{Sec:concl}

Heavy neutral leptons are one of the best motivated BSM candidates and, when heavy, high-energy colliders provide an optimal laboratory for their study. They can be produced resonantly, leading to the most commonly searched channels, but only if they are kinematically accessible. At the LHC, for instance, this happens for masses up to the TeV. In contrast, t-channel searches provide a path to extend collider probes to heavier masses. 

In this work we have revisited neutrino-mediated t-channels, showing the importance of coherently adding both light --disregarded in some previous literature-- and heavy neutrino contributions to properly reproduce the predictions of the model. We focused mostly on hadronic colliders, but our results also translate to leptonic ones. We concluded that, although LNV t-channels are not the most promising signals to probe the type-I seesaw mechanism, LNC t-channels represent very promising searches, specifically the LFV ones.

Taking the $e\mu jj$ signal as case of study, we have shown that the LHC could improve current collider bounds from resonant searches for masses above $m_N\gtrsim800$~GeV. Furthermore, this search is sensitive to low-scale seesaw pseudo-Dirac HNLs, which escape the standard LNV searches. The same search can be applied to $\tau\ell j j$ signals, although we leave this study for future work. 

Our analysis shows that the LHC could reach mixings of $V^2\leq \mathcal O(10^{-2}-10^{-1})$ at current runs or in the high-luminosity phase for masses in the few TeV range. Interestingly, for very heavy masses, these sensitivities could actually be understood as searches for the dimension 6 operator generated after integrating out the HNLs, with $\eta_{e\mu}\lesssim \mathcal O(10^{-2})$.
Unfortunately, such sensitivities are far from our current best limits of $\eta_{e\mu}\leq \mathcal O(10^{-5})$ from precision data~\cite{Blennow:2023mqx}, in particular from $\mu\to e\gamma$ transitions~\cite{MEGII:2025gzr}.
Still, given our current lack of indications for new physics, it is important to explore every possible scenario and, in this context, the LHC provides a complementary probe of the heavy HNL hypothesis. 

%All in all, we have shown that t-channels are indeed a promising path to extend the reach of colliders to heavier masses, but special attention must be taken in order to consistently consider all possible contributions. In particular, we conclude that LFV t-channels are an appealing low-background search which is worth pursuing at current and future LHC runs. 

In summary, we have shown that HNL-mediated t-channel processes provide a promising avenue to extend the mass reach of collider searches beyond the kinematic limit of direct production, provided that all contributions required by the seesaw framework are consistently included. In this context, LFV t-channel signatures emerge as particularly promising, offering a clean, low-background probe of low-scale seesaw scenarios at the LHC and motivating dedicated searches in current and future runs.

 \medskip
 \paragraph{Acknowledgments.}
 We would like to thank Enrique Fern\'andez-Mart\'inez for his valuable insight, both in the early and later stages of this work. CGG would like to express her gratitude to Nicol\'as Escudero and Javier Quilis, whose wisdom and support continue to resonate from their days as physicists right up to the present day. The work of CGG and XM is funded by the Italian Ministry of Universities and Research (MUR) and the European Union - Next Generation EU, Missione~4 Componente 1 CUP J33C24003210006 - NEWTRINOS, and by the Italian INFN program on Theoretical Astroparticle Physics (TAsP). The work of DNT is supported by the Alexander von Humboldt foundation.

%%%%%%%%%%%%%%%%%%%%%%%%%%%%%%%%%%%%%%%

% \appendix

% \section{Some appendix}\label{App:A}

% More details...

%%%%%%%%%%%%%%%%%%%%%%%%%%%%%%%%%%%%%%%%

\bibliographystyle{JHEP} 
\bibliography{biblio}% Produces the bibliography via BibTeX.

\end{document}